\documentstyle[12pt]{article}
\def\a{\alpha}
\def\b{\beta}

\def\d{\delta}

\def\g{\gamma}

\def\l{\lambda}
\def\m{\mu}
\def\n{\nu}

\def\p{\pi}

\def\r{\rho}

\def\G{\Gamma}

\def\cl{{\cal L}}
\def\cm{{\cal M}}

\def\rt{\rightarrow}

\def\bar#1{\overline{#1}}

\def\Hat#1{\rlap{\kern.10em$\widehat{\phantom G}$}#1}
\def\HAt#1{\rlap{\kern.05em$\widehat{\phantom G}$}#1}

\def\cap#1{\rlap{\kern.1em$\widehat{\phantom{G\vrule height.8em}}$}#1{}}
\def\Cap#1{\rlap{\kern.05em$\widehat{\phantom{G\vrule height.8em}}$}#1{}}

\catcode`\@=11
\long\def\@makefntext#1{ %\parindent 1em
\protect\noindent \hbox to 3.2pt {\hskip-.9pt
$^{{\ninerm\@thefnmark}}$\hfil}#1\hfill} %can be used

\def\thefootnote{\fnsymbol{footnote}}
 \def\@makefnmark{\hbox to 0pt{$^{\@thefnmark}$\hss}}  %original

\def\ps@myheadings{\let\@mkboth\@gobbletwo
\def\@oddhead{\hbox{} %\sl
\rightmark\hfil\ninerm\thepage}
\def\@oddfoot{}\def\@evenhead{\ninerm\thepage\hfil %\sl
\leftmark\hbox{}}\def\@evenfoot{}
\def\sectionmark##1{}\def\subsectionmark##1{}}

\begin{document}

%----------------------------PROCSLA.STY---------------------------------------
\newcommand{\symbolfootnote}{\renewcommand{\thefootnote}
        {\fnsymbol{footnote}}}
\renewcommand{\thefootnote}{\fnsymbol{footnote}}
\newcommand{\alphfootnote}
        {\setcounter{footnote}{0}
         \renewcommand{\thefootnote}{\sevenrm\alph{footnote}}}

%------------------------------------------------------------------------------
%NEW DEFINED SECTION COMMANDS
\newcounter{sectionc}\newcounter{subsectionc}\newcounter{subsubsectionc}
\renewcommand{\section}[1] {\vspace{0.6cm}\addtocounter{sectionc}{1}
\setcounter{subsectionc}{0}\setcounter{subsubsectionc}{0}\noindent
        {\bf\thesectionc. #1}\par\vspace{0.4cm}}
\renewcommand{\subsection}[1] {\vspace{0.6cm}\addtocounter{subsectionc}{1}
        \setcounter{subsubsectionc}{0}\noindent
        {\it\thesectionc.\thesubsectionc. #1}\par\vspace{0.4cm}}
\renewcommand{\subsubsection}[1]
{\vspace{0.6cm}\addtocounter{subsubsectionc}{1}
        \noindent {\rm\thesectionc.\thesubsectionc.\thesubsubsectionc.
        #1}\par\vspace{0.4cm}}
\newcommand{\nonumsection}[1] {\vspace{0.6cm}\noindent{\bf #1}
        \par\vspace{0.4cm}}

%NEW MACRO TO HANDLE APPENDICES
%------------------------------------------------------------------------------
%MARCO FOR ABSTRACT BLOCK
\def\abstracts#1{{
        \centering{\begin{minipage}{30pc}\tenrm\baselineskip=12pt\noindent
        \centerline{\tenrm ABSTRACT}\vspace{0.3cm}
        \parindent=0pt #1
        \end{minipage} }\par}}

%------------------------------------------------------------------------------

%------------------------------------------------------------------------------

%LIST ENVIRONMENTS

%------------------------------------------------------------------------------
%------------------------------------------------------------------------------
%------------------------------------------------------------------------------
\font\twelvebf=cmbx10 scaled\magstep 1
\font\twelverm=cmr10 scaled\magstep 1
\font\twelveit=cmti10 scaled\magstep 1
\font\elevenbfit=cmbxti10 scaled\magstephalf
\font\elevenbf=cmbx10 scaled\magstephalf
\font\elevenrm=cmr10 scaled\magstephalf
\font\elevenit=cmti10 scaled\magstephalf
\font\bfit=cmbxti10
\font\tenbf=cmbx10
\font\tenrm=cmr10
\font\tenit=cmti10
\font\ninebf=cmbx9
\font\ninerm=cmr9
\font\nineit=cmti9
\font\eightbf=cmbx8
\font\eightrm=cmr8
\font\eightit=cmti8

%----------------------START OF DATA FILE------------------------------

\centerline{\tenbf POSSIBLE EXTENSION OF THE CHIRAL}
\baselineskip=22pt
\centerline{\tenbf PERTURBATION THEORY PROGRAM}
\vspace{0.8cm}
\centerline{\tenrm J. SCHECHTER}
\baselineskip=13pt
\centerline{\tenit Department of Physics, Syracuse University,}
\baselineskip=12pt
\centerline{\tenit Syracuse, NY 13244-1130, USA}
\vspace{0.9cm}
\abstracts{ After a brief discussion of how chiral dynamics has evolved from
the
``universal V-A theory of weak interactions'', we present some evidence that
symmetry breaking for the vector meson multiplet is not simpler than but rather
analogous to that for the pseudoscalar multiplet.  This provides a motivation
for speculating on how to extend in a systematic way the chiral perturbation
theory program to include vectors.
}
\vfill
\twelverm   %modified by CLee 23/07/93
\baselineskip=14pt
\section {Introduction}
I would like to dedicate this paper to the memory of Professor Robert Marshak.
As an ex-graduate student at Rochester I am grateful to him for establishing an
intellectually stimulating and supportive Particle Physics group there.  His
enthusiasm for research and down-to-earth attitude were much appreciated by all
of us.

Out of a career filled with many achievements in physics, Professor Marshak's
chief one was the deduction, together with E.C.G. Sudarshan, of the ``V-A''
form of the weak interaction.$^1$  On the one hand, this theory provided a
basis for understanding a  wealth of experimental data.  Equally important, it
indicated that the relevant degrees of freedom of the  observed ``material''
particles in nature were not the Dirac spinor fields but rather their left and
right chiral projections.  The left projections appeared in the weak
interactions while both left and right were needed for the strong and
electromagnetic interactions.

The most evident application of this idea to the strong interactions of the low
lying - i.e. pseudoscalar meson - states requires us to increase the size of
the
``flavor'' symmetry multiplet.  Instead of treating the $0^-$ mesons as
belonging to a $3\times 3$ matrix $\Phi$ which transforms as $\Phi \rt U \Phi
U^{-1}$,
$U^{\dagger}U=1$ under the approximate symmetry group of the light quarks, we
associate a $3\times 3$ matrix S of scalar mesons with $\Phi$ to form an object
$M=S+i\Phi$ which transforms as $M \rt U_LMU_R^{\dagger}$ under separate left
projected and right projected unitary groups.

Some consequences of this ``linear'' chiral symmetry approach were discussed by
Prof. Marshak and collaborators in Ref. 2.  However it turns out that another
important physical ingredient is required.  The vacuum of the strong
interaction theory is not (even in the limit of massless light quarks)
invariant under the separate $U_L$ and $U_R$ transformations, although the
Lagrangian is invariant.  As Nambu$^3$ explained to the world, this implies
that the pseudoscalar and scalar masses are drastically split from each other,
the
pseudoscalar masses being forced to become zero in the massless quark limit.
This situation can be  neatly handled by making the polar decomposition $M=HU$,
$H=H^{\dagger}, U^{\dagger}=U^{-1}$ and `` freezing out'' the scalar field part
$H$ by setting $H=(F_\pi/2)1$.  The freezing out corresponds to sending the
scalar masses to infinity while $F_\p \simeq$ 0.132 GeV is the  pion decay
constant.  The name of this basic constant of strong interaction physics
betrays its origin in the weak interactions.  Note that $U\rt U_L U
U_R^{\dagger}$ under chiral transformations.  $U$ is a function of the ordinary
pseudoscalar field matrix $\phi$ (which is simply related to $\Phi$ and $S$ by
a ``point transformation'') and one may conveniently set$^4$ $U=\exp
(2i\phi/F_\pi)$. $\phi$ evidently behaves non-linearly under chiral
transformations.  Having isolated the proper degrees of freedom it remains only
to note that the simplest invariant Lagrangian density formed with $U$,
$$
\cl = -\frac {F^{2}_{\pi}}{8} ~\rm {Tr}~ (\partial_\m U \partial_\m
U^{\dagger})+... ~, \eqno (1.1)
$$
\noindent provides an accurate representation of QCD at very low energies
(where
perturbation  theory fails).  Indeed, many of the ``current algebra'' theorems,
carefully discussed in the treatise ``Theory of Weak Interactions in Particle
Physics''$^5$ can be obtained using (1.1) in a simple way.

It seems remarkable that, although many years have passed since the
introduction of the ``V-A'' theory, chiral symmetry is still an extremely
active field of research.  In the following, I shall very briefly describe some
of the progress in the field and shall present some speculations related to
further improvement of the approach.

\section {Going Beyond Very Low Energies}
Modifying the simple model (1.1) in an attempt to describe QCD over the full
traditional low energy range (say up to about 1 GeV) has been the task of a
generation.  And it has still not been definitively accomplished.  It is, of
course, necessary to recognize that the chiral symmetry is broken by the light
quark mass terms$^6$ in the fundamental QCD Lagrangian,
$$
\cl_{\rm mass} =- \hat {m} \bar {q} \cm q, \eqno (2.1)
$$
\noindent where $q$ is the column vector of up, down and strange quark fields,
 $\hat m= (m_u+m_d)/2$ and $\cm$ is a dimensionless, diagonal matrix which can
be expanded as follows:
$$
\cm = y \l_3 + T + x S , \eqno (2.2)
$$
\noindent with $\l_3= ~\rm {diag} ~(1,-1,0),T=~\rm {diag} (1,1,0)$ and $S=
\rm {diag} (0,0,1)$.  $x$ and $y$ are the quark mass ratios:
$$
x= \frac {m_s}{\hat{m}},~y=-\frac {1}{2} \left ( \frac {m_{d}-m_{u}}
{\hat{m}}\right ).       \eqno (2.3)
$$
\noindent The minimal term at the effective Lagrangian level which can mock up
(2.1) is:
$$
\cl_{SB}= \delta^{\prime} ~\rm {Tr}~ [\cm (U+U^{\dagger}-2)], \eqno(2.4)
$$
\noindent where $\d'$ is a numerical constant.  Fitting (1.1) $+$ (2.4) to the
experimental pseudoscalar mass spectrum yields the standard determination of
the quark mass ratios $x$ and $y$.

Now, there are two (in principle) straightforward ways in which one might
extend up in energy the description of low energy physics obtained by using
(1.1) $+$
(2.4) at tree level:
\vglue 0.3cm
\noindent {\it i. Including loop diagrams}$^7$
\vglue 1pt

First compute the one loop corrections
using (1.1) $+$ (2.4)  and keep terms quartic in the momenta.  To eliminate the
divergences  add chiral invariant ``counter-terms'' quartic in derivatives.
For
this purpose count each power of $\cm$ as two derivatives (The terms involving
$\cm$ will be chiral covariant rather than invariant).  This scheme can be
continued to higher orders in momenta.  However since the starting Lagrangian
is non-renormalizable there will be an infinite number of counterterms.  This
is not a worry in practice as we only expect to use the method up to a low
finite order.
\vglue 0.3cm
\noindent {\it ii. Including additional physical particles}
\vglue 1pt

Even without considering any detailed models it is obvious that if we want to
have an effective theory valid up to some energy we should include the physical
particles whose masses lie in that energy range.    How else would we get the
right poles in the S-matrix at tree level?  This point of view is buttressed by
the Veneziano model$^8$ which shows how to get good high energy behavior by
adding
pole contributions from a realistic looking (infinite) set of particles.  It
also is supported by the leading large $N_c$ approximation$^9$ to QCD in which
one should keep only the tree diagrams involving all the physical mesonic
states.
The baryonic states can be obtained from this effective meson Lagrangian as
Skyrme solitons.$^{10}$

Which of these two approaches is superior, or is that even the right question
to ask?  The first approach, which has been systematized by Gasser and
Leutwyler$^7$ is known as ``chiral perturbation theory'' (CPT).  In practice
it essentially amounts to making a complete list of chiral invariant and
covariant counterterms, each with an unknown coupling constant.  There are
about ten of these at the quartic derivative order. With a convenient choice of
the renormalization point, the contributions from the loop diagrams themselves
(the
``chiral logs'') are typically negligible.  It seems that the CPT approach is
both solid and useful for improving the description of (1.1) $+$ (2.4) at very
low energies, say up to about 500 MeV in $\pi-\pi$ scattering.  But going
beyond this region forces us to face the enormous peak representing the $\r$
meson.  It is hard to avoid including it ( and all its $SU(3)$  partners) if we
want a realistic description.  In fact it has been found$^{11}$ that many of
the
values of the counterterms can be numerically understood just with vector
meson pole dominance.  These arguments strongly suggest the  suitability of a
model of type ii.  Should we then give up the CPT program?   Here, we would
like to argue for a model combining the two approaches.

Immediately, there may be a number of objections.  First it seems to be
discouraging to have to include every particle multiplet with the same kind of
microscopic detail that has been applied to the pseudoscalar multiplet.  In
response, we may note that a natural continuation of the present CPT program
would be simply to include at first  just the vector meson multiplet.  This
provides a ``clean break'' in the sense of retaining just the low lying S-wave
quark anti-quark states in the model.  It would provide coverage of the region
up to around 1 GeV.  The lessons learned in such a generalization may show us
how to economically include the still more massive states.

Another  possible objection is related to the general feeling that, since the
low lying pseudoscalars are approximate Nambu-Goldstone bosons, they should be
treated differently from the other multiplets.  This apparent objection would
appear to be strengthened by the folk wisdom that while one can explain the
properties of a ``normal'' multiplet like the vectors with a minimal deviation
from $SU(3)$ symmetry, much more elaboration, via the inclusion of many
arcane symmetry breaking terms on the CPT list of counterterms, is required for
the
pseudoscalar multiplet.  Here, we would like to point out that this folk
wisdom
does not seem to hold.  In a recent paper,$^{12}$ which should be consulted for
more explanations and references, it is shown that exactly analogous symmetry
breaking terms are required for both the vector and pseudoscalar multiplets at
the Okubo-Zweig-Iizuka (OZI)  rule  conserving level.  This suggests that all
multiplets be treated in the same way.  The (approximate) spontaneous breakdown
of chiral symmetry is certainly a crucial feature but it would appear to
affect {\em every} multiplet, presumably via the ``pion cloud'' intrinsic to
each
particle.  Of course, the vector and higher multiplets have non-zero masses in
the chiral limit.  One might imagine that the appropriate ``large'' scale with
which to compare the effects of the perturbation $(\hat{m}\cm)$ is
$(\a')^{-1/2}\simeq $
1.06 GeV, $~\a'$ being the universal Regge slope parameter.  The best choice of
renormalization point for the loop diagrams requires investigation.

In Ref. 12 the vector meson nonet field $\r_\m(x)$ is introduced, for
convenience, in terms of auxiliary, linearly transforming ``gauge fields''
$A^L_\m$ and $A^R_\m$ by$^{13}$:
$$
A^L_\m = \xi \rho_\m \xi^{\dagger} +
\frac {i}{\tilde{g}} \xi \partial_\m\xi^{\dagger},~
A^R_\m=\xi^{\dagger}\rho_\m\xi +
\frac {i}{\tilde{g}} \xi^{\dagger}\partial_\m \xi,   \eqno(2.5)
$$
where $\xi =U^{1/2}$ and $\tilde{g}$ is related to the $\rho \phi \phi $
coupling constant.  $\rho_\m$ transforms non-linearly in this description,
which corresponds to eliminating the axial vector mesons in analogy to the
elimination of the scalar mesons  which led to the non
linearly transforming pseudoscalar multiplet inside $U$.  Note that both the
scalars and axials are $P$-wave $q\bar{q}$ bound states so this truncation is
conceptually consistent.

Now we can play the CPT game, constructing all chiral invariants and covariants
up to a certain order in derivatives. Both vector and pseudoscalar fields would
be included.
This should necessitate  readjusting the
coefficients of those terms containing only pseudoscalars which were dominated
by vector meson exchange.  To start to explore this rather complicated scheme
we will, first of all, specialize to the symmetry breaking terms and try to fit
all the mass differences, including those which are isospin violating.  We will
also fit the  meson decay constants and $V\rt \phi \phi$ decay widths.  As a
physical approximation we shall demand that (with one significant exception)
the
symmetry breaking terms be single traces in flavor space and that all
field matrices represent nonets.  This is Okubo's form$^{14}$ of the OZI rule
and
appears to be  respected by the existing CPT fit for the  pseudoscalar only
symmetry  breakers.  The final approximation is to neglect the ``chiral logs''.
This also works in the existing CPT  fits.  Of course, these approximations can
be relaxed in the future.

Then the symmetry breaking terms which conserve the OZI rule are taken to be
(up to quartic order in derivatives):
$$
\cl_{SB}=~\rm {Tr}~ \{\cm [\d'(U+U^{\dagger}-2) + \a'(A^L_\m
UA^R_\m + A^R_\m U^{\dagger}A^L_\m)
$$
$$
+\b' (\partial_\m U  \partial_\m  U^{\dagger} U +
U^{\dagger}\partial_\m U\partial_\m U^{\dagger}) +
\g' (F^L_{\m\n}UF^{R}_{\m\n} + F^R_{\m\n}U^{\dagger}F_{\m\n}^{L})]
$$
$$
+ \l'^{2} [\cm U^{\dagger}\cm U^{\dagger}+\cm U \cm U -2 \cm^2]+\m'
(A^L_\m \cm A^R_\m \cm)\} \eqno (2.6)
$$
\noindent where $F^{L,R}_{\m\n}= \partial_\m A^{L,R}_{\n}
-\partial_{\n}
A^{L,R}_{\m} - i \tilde {g} [ A^{L,R}_\m, ~A^{L,R}_\n]$ while
$\a',\b',\g',\d',\l^{\prime 2},\m^{\prime}$
are constants to be determined.
Notice that there are three analogous vector terms and  pseudoscalar terms.
Physically, each multiplet has a non-derivative and a derivative type symmetry
breaker proportional to $\cm$ as well as a non-derivative term proportional to
$\cm^2$.  It turns out that they are {\em all} required to fit the pseudoscalar
and vector particle properties mentioned above.

As explained in section III of Ref. 12 it is convenient to determine a suitable
number of the physical quantities while holding the quark mass ratio $x$
constant.  Then there are three predictictions for each value of $x$.  A best
fit
is obtained for the quark mass ratios:
$$
x=37~,~~~y=-0.36.\eqno(2.7)
$$
If the $\g'$ term for the vectors were not present it would be very difficult
to get reasonable predictions for the non-electromagnetic  part of
$m(K^{*0})-m (K^{*+})$ and for the width ratio $\G(K^*)/\G(\r)$.  The effect of
this derivative type symmetry breaking term is to introduce non-trivial wave
function renormalizations for the $K^*$ and $\phi$ particles.  Similarly, the
$\m'$ term improves the predictions of the mass and width of the $\phi$ meson
in the  present framework.

For the pseudoscalars, it is well known that we can not restrict ourselves to
just the OZI rule conserving terms.  The needed extra terms are discussed in
sections II(c) and IV of Ref. 12.  There it is shown that the minimal
Lagrangian which can solve the $U(1)$ problem (with the aid of an auxiliary
glueball
field) can be modified by the addition of suitable symmetry breaking terms to
give
a reasonable description of  the $\eta-\eta'$ system.  In particular, the old
problem$^{15}$  of too small $\eta$ mass, which has more recently attracted
some
attention,$^{16}$ was solved.  A consequence of this discussion is the presence
of the OZI rule violating term $\propto$
$$
\{\rm Tr~ [\cm (U-U^{\dagger})]\}^2. \eqno(2.8)
$$

On the other hand, there is no special reason to include a sizeable term of the
type
$$
\{\rm Tr~ [\cm (U+U^{\dagger})]\}^2. \eqno(2.9)
$$

The inclusion of (2.8) but not (2.9), which is very natural in the approach
mentioned above, amounts to a practical resolution of the Kaplan-Manohar
ambiguity.$^{17}$  Some discussion is given in section VII(c) of Ref. 12.  We
also have no special reason to include the pseudoscalar OZI rule violating
term
of the type $\propto$
$$
\rm {Tr}~(\partial_\m U \partial_\m U^{\dagger}) ~Tr~[\cm (U+U^{\dagger})].
\eqno (2.10)
$$

This term is found to be very small in the usual CPT fit with pseudoscalars
only. We have now accounted for all the terms depending upon $\cm$ on the CPT
list for pseudoscalars.  The only important OZI rule violating one was, as
expected, the one
associated with the $U(1)$ problem and the $\eta-\eta '$ system. Thus our
identification of the most important symmetry breaking terms
seems justified for our purpose of making an appraisal of the validity of this
model.

Incidentally, we remark that the quark mass ratios in (2.7) are somewhat
different from the usual$^{18}$ ones $x=25.0 \pm 2.5$, $y=-0.28 \pm 0.03.$
With our
determination and  a {\em choice} $ m_s$(1 GeV)=0.175 GeV we would have
$m_u$=3.2 MeV and $m_d$=7.9 MeV.

To sum up, we have presented evidence that, apart from the special terms needed
to solve the $U(1)$ problem, the treatment of symmetry breaking in the
pseudoscalar and vector  multiplets involves exactly analogous terms.  The
implication is that chiral perturbation theory might be extensible, as
outlined, to
the vectors too in order to model low energy QCD up to around 1 GeV.  Certainly
a large number of processes, including loop contributions, must be examined to
fully test this idea.  But the first step is to make a preliminary calculation
by picking out the terms expected to be most important and working at tree
level.  The treatment of symmetry breaking discussed here in that manner
provides an optimistic sign.

\section {Acknowledgments}
I would like to thank Anand Subbaraman and Herbert Weigel for a pleasant
collaboration.

\section {References}
\begin{enumerate}

\item  E.C.G. Sudarshan and R.E. Marshak, {\it Proceedings of the  Padua
conference on mesons and recently discovered particles}, p. V-14 (1957).  See
also R.P. Feynman and M. Gell-Mann, {\it Phys. Rev.} {\bf 109}, 193 (1958) and
J.J. Sakurai, {\it Nuovo Cimento} {\bf 7}, 649 (1958).

\item R.E. Marshak, N. Mukunda and S. Okubo, {\it Phys. Rev} {\bf 137} B698
(1965).

\item Y. Nambu in {\it Group Theoretical Concepts and Methods in Elementary
Particle Physics}, edited by F. G\"{u}rsey (Gordon and Breach, New York, 1964).

\item K. Nishijima, {\it Nuovo Cimento}, {\bf 11} 698 (1959); F. G\"{u}rsey,
{\it
Ann. Phys.} (N.Y.) {\bf 12} 91 (1961); J. Cronin, {\it Phys. Rev.} {\bf 161},
1483 (1967).

\item R.E. Marshak, Riazuddin and C.P. Ryan, {\it Theory of Weak Interactions
in
Particle Physics}, Wiley-Interscience (1969).

\item M. Gell-Mann, {\it Phys. Rev.} {\bf 125}, 1067 (1962).

\item L.-F. Li and H. Pagels, {\it Phys. Rev. Lett.} {\bf 26}, 1204 (1971);
S. Weinberg, {\it Physica} {\bf 96A}, 327 (1979); J. Gasser and L. Leutwyler,
{\it Nucl. Phys.} {\bf 250}, 465 (1985).

\item G. Veneziano, {\it Nuovo Cimento} {\bf 57A}, 190 (1968); See also L. Van
Hove, {\it Phys. Lett.} {\bf 24B}, 183 (1967).

\item G. 't Hooft, {\it Nucl. Phys.} {\bf B72}, 461 (1974); E. Witten {\it
ibid}, {\bf B160}, 57 (1979).

\item T. Skyrme, {\it Proc. Roy. Soc.} {\bf A262}, 237 (1961).

\item J. Donoghue, C. Ramirez and G. Valencia, {\it Phys. Rev.} {\bf D39}, 1947
(1989);  G. Ecker, J. Gasser, A.Pich and E. de Rafael, {\it Nucl. Phys.} {\bf
B221}, 311 (1989).

\item J. Schechter, A. Subbaraman and H. Weigel, {\it Phys. Rev.} {\bf D48},
339 (1993).

\item \"{O}. Kaymakcalan and J. Schechter, {\it Phys. Rev.} {\bf D31}, 1109
(1985).  An alternate equivalent approach is given by T. Fujiwara, T. Kugo, H.
Terao, S. Uehara and K. Yamawaki, {\it Prog. Theor. Phys.} {\bf 73}, 926
(1985).  The equivalence is demonstrated in P. Jain, R. Johnson, U. Meissner,
N.W. Park and J. Schechter, {\it Phys. Rev.} {\bf D37}, 3252 (1988).

\item S. Okubo, {\it Phys. Lett.} {\bf 5}, 165 (1963).

\item See Fig. 2 of V. Mirelli and J.
Schechter, {\it Phys. Rev.} {\bf D15} 1361 (1977).

\item H. Georgi, {\it Phys. Rev.} {\bf D49}, 1666 (1994); S. Peris , CERN
preprint TH
7109/93.

\item D. Kaplan and A.  Manohar, {\it Phys. Rev. Lett.} {\bf 56}, 2004 (1986).

\item J. Gasser and H. Leutwyler, {\it Phys. Rep.} {\bf 87}, 77 (1982).

\end{enumerate}
\end{document}